# Recognition of Nanoparticle and Microdroplet Parameters in Levitating Microdroplets of Suspension by Speckle Image Analysis Using Convolutional Neural Networks

*Yaroslav Shopa [1], Kwasi Nyandey [2,3], Daniel Jakubczyk [3,*]*

[1] Faculty of Mathematics and Natural Sciences. School of Exact Sciences, Cardinal Stefan Wyszynski University in Warsaw, Warsaw 01-815, Poland
[2] Department of Physics, School of Physical Sciences, Laser and Fibre Optics Centre, College of Agriculture and Natural Sciences, University of Cape Coast, Cape Coast, Ghana
[3] Institute of Physics, Polish Academy of Sciences, Warsaw 02-668, Poland

**Corresponding author: jakub@ifpan.edu.pl*

## Abstract

Retrieving suspension parameters from light scattered by a single suspension microdroplet is a difficult inverse problem because the recorded optical pattern is shaped by droplet morphology and by perturbations of the internal optical field introduced by the dispersed phase. Here we demonstrate experimental, data-driven recognition of microdroplet and nanoparticle-suspension parameters from laser speckle images recorded for individual levitated suspension microdroplets. Slowly evaporating $TiO_2$ nanoparticle suspensions in diethylene glycol were confined in a linear electrodynamic quadrupole trap, and their speckle patterns were classified using a convolutional neural network. The network was trained and tested to recognize droplet diameter, nanoparticle mass concentration, and nanoparticle diameter, first separately and then in simultaneous two- and three-parameter tasks. Droplet diameter was the most robustly recognized quantity, with an estimated accuracy better than approximately 6% for independent test data. Nanoparticle concentration required sufficiently separated classes. Nanoparticle-diameter classes were also recognized, although particle-size effects were coupled to changes in particle number at fixed mass concentration. Simultaneous three-parameter classification across 27 classes was demonstrated. As an exploratory out-of-distribution test, a network trained on single-component suspensions was applied to dual-component suspensions containing two nanoparticle diameters. The network retained meaningful predictive capability, especially at higher concentration, although robust decomposition of mixed nanoparticle populations was not achieved. These results show that speckle images of single levitated suspension microdroplets contain sufficient information for CNN-based multi-parameter recognition and support further development of speckle-based optical diagnostics for complex microdroplet and aerosol-like systems.

## Introduction

The optical response of a suspension microdroplet is determined jointly by the properties of the dispersed phase and by the fact that the droplet itself is a finite dielectric object. In contrast to a bulk suspension, a microdroplet supports optical morphology-dependent resonances and therefore modifies the optical field interacting with the suspended phase [1]; see also [2]. At the same time, the internal field of an equivalent ideal spherical resonator is modified by the presence of the suspended particles. As a consequence, the total scattered field need not reflect only the properties of the suspended-phase particles, but is strongly shaped by the droplet-scale effective field distribution.

Under coherent (laser) illumination, this complex scattered field produces a spatial laser-speckle intensity distribution, recorded by the camera as a two-dimensional pattern. Laser speckle is particularly sensitive to slight changes in the structure of the scattering object because it arises from the interference of many partial wave contributions. Its spatial structure can encode subtle information about the droplet geometry, the effective optical properties of the medium, and the presence, distribution, and properties of suspended nanoparticles [3]; see also [4]. However, the relevant information is distributed over a complex stochastic texture rather than contained in a small number of directly identifiable observables; see, e.g., [5]. Thus, the retrieval of suspension and droplet parameters from recorded speckle patterns is a non-trivial inverse problem.

Speckle-like and highly structured two-dimensional scattering patterns from particles, aggregates and droplet-derived systems have been studied previously. Two-dimensional angular optical scattering (TAOS) has been used to characterize airborne microparticles and aggregates, and multispectral TAOS measurements have been combined with morphological descriptors for preliminary particle classification. In particular, Holler et al. demonstrated simultaneous two-colour TAOS acquisition from airborne particulates and showed that descriptors extracted from the scattering images could support classification of aerosol types [6]. Related studies have shown that scattering patterns from complex particles and aggregates contain morphology-dependent information that can be exploited by automated analysis methods. More recently, supervised machine-learning methods have been applied to simulated multispectral TAOS patterns of particle aggregates [7], and convolutional-neural-network-based approaches have been used to analyse complex scattering or speckle-like particle images [8,9].

Convolutional neural networks (CNNs) are well suited for extracting information encoded in spatially distributed scattering patterns. They can learn discriminative image features directly from experimental data and combine them hierarchically into task-relevant representations [10–12]. In image-analysis tasks where the informative content is texture-like and spatially distributed, CNNs are often more effective than manually designed descriptors because they do not require prior identification of which image features are physically most relevant (compare [13]). Indeed, speckle-pattern analysis combined with machine learning has been explored for liquid and suspension characterization. Examples include the identification of pharmaceutical suspensions from speckle pattern images using machine-learning classifiers [14], and raw-milk classification using both feature-based machine learning and image-based deep learning [13]. These studies support the broader view that speckle images can act as informative fingerprints of complex and/or turbid liquid samples.

In our previous work, we showed that CNN-based classification of laser speckle images can identify monodisperse nanoparticle suspensions in water, including suspended-particle type and concentration [15]. We later extended this approach to more complex real-world suspensions, exemplified by cow milk, for which the fat content was classified from speckle images [16]. We also showed that a CNN-based classification task can be formulated to follow the diameter evolution of levitated evaporating microdroplets of pure liquids, liquid mixtures and suspensions [17].

These earlier studies, however, do not directly address the inverse problem considered here: whether experimentally recorded speckle images from a single levitated suspension microdroplet can be used to recognize, in a data-driven way, both droplet parameters and parameters of the dispersed phase. This distinction is important because a suspension microdroplet is not simply a small volume of bulk suspension. Consequently, information about the target parameters – droplet diameter, nanoparticle concentration, and nanoparticle diameter – is encoded jointly in a complex speckle texture.

The present work extends our previous speckle-CNN studies [15–17] to slowly evaporating levitated microdroplets of monodisperse $TiO_2$ nanoparticle suspensions in diethylene glycol. The microdroplets were confined in a linear electrodynamic quadrupole trap of the type used in our earlier studies of suspension microdroplets [18]. The specific objective is to test experimentally whether CNN-based speckle-image analysis can recognize selected microdroplet and suspension parameters from images recorded for individual levitated droplets. We fo-cus on three quantities: droplet diameter, nanoparticle mass concentration and nanoparticle diameter. These are first considered in separate classification tasks and then in simultaneous two-parameter and three-parameter tasks.

The main contribution of this work is the demonstration that experimental speckle images from single levitated suspension microdroplets can support multi-parameter classification of both droplet and dispersed-phase properties. In this proof-of-principle study, the main classification experiments were intentionally restricted to single-component, monodisperse $TiO_2$ nanoparticle suspensions. This model system is experimentally manageable, provides sufficient optical contrast and allows the classification results to be interpreted more clearly than would be possible for a fully polydisperse or chemically complex aerosol. As an additional exploratory step, we also test how a network trained on single-component suspensions responds to dual-component suspensions containing two nanoparticle-size fractions. This test is treated as a probe of generalization beyond the training distribution, not as a complete solution to the decomposition of mixed nanoparticle populations.

## Experimental setup and procedures

The investigated microdroplets were maintained in a custom-built linear electrodynamic quadrupole trap (LEQT), described previously in [18,19] and shown schematically in Figure 1. Electrodynamic levitation enables non-contact observation of individual droplets while providing optical access for imaging and light-scattering measurements [20–23]. In aerosol and droplet physics, such methods are widely used to investigate evaporation, hygroscopicity, morphology, phase transitions, and other physicochemical properties of isolated particles. Here, it was used to record laser speckle images from slowly evaporating microdroplets of nanoparticle suspensions. The overall image-analysis workflow and CNN architecture are summarized in Figure 2 and described below.

For each experiment, a target initial suspension concentration was specified and the suspension was prepared to the corresponding concentration. Because the microdroplets evaporated slowly, the change in microdroplet diameter during the acquisition of a single speckle-image series was comparable to the accuracy of diameter measurement (~0.1%) and could therefore be neglected. Similarly, the average suspension concentration during such an acquisition interval could be regarded as effectively constant, with a change of less than 0.35%. Depending on the NP diameter and concentration and microdroplet diameter, microdroplets contained from ~$10^3$ to ~$10^7$ NPs (for amorphous $TiO_2$ density of 2.2 g/cm$^3$ declared by the manufacturer [24]). The average distance between NPs in the droplet volume varied from ~0.2 μm to ~3 μm, which, again depending on the NP diameter, was in the range from ~2 to ~6 NP diameters.

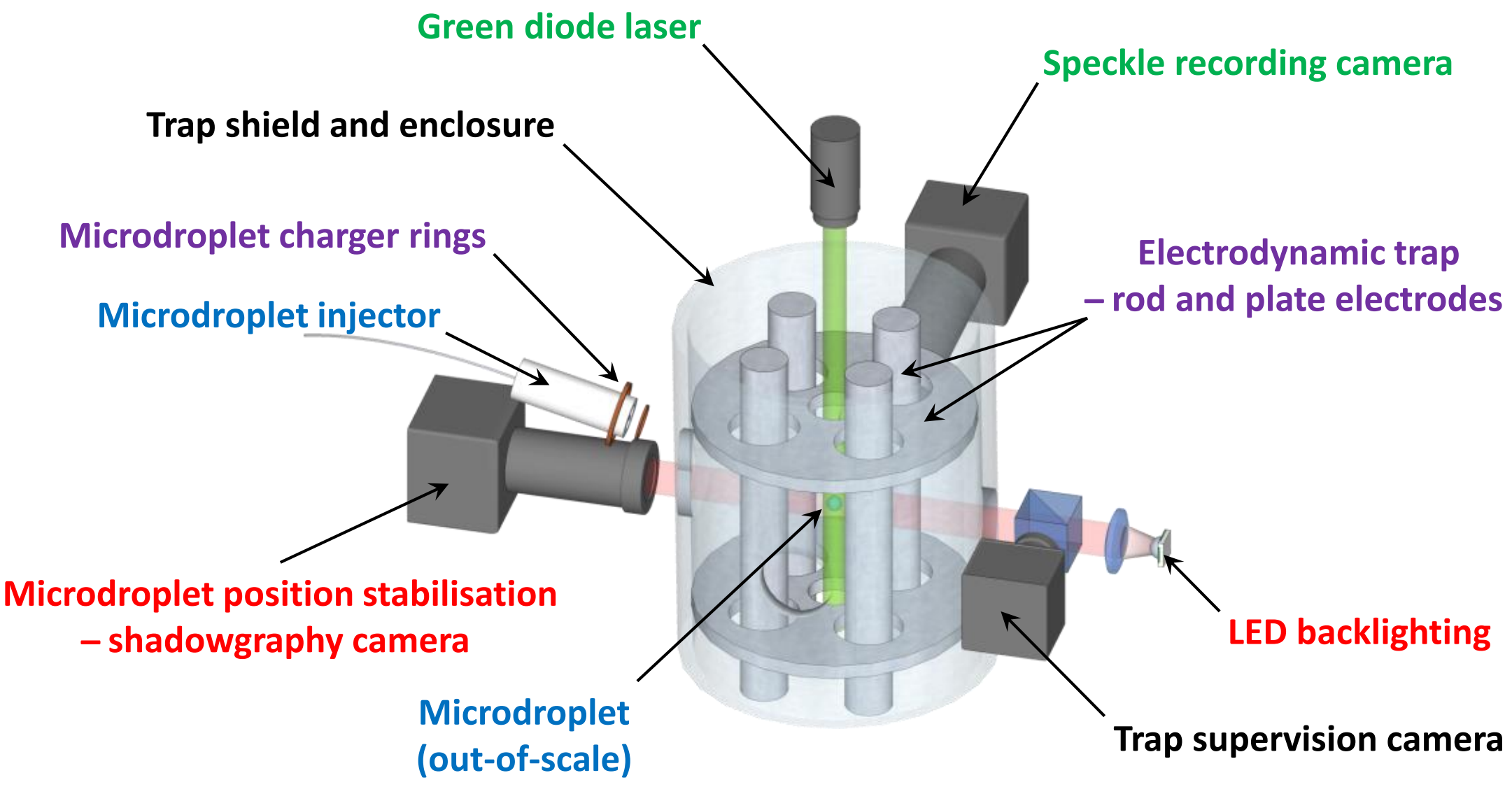

Figure 1 Simplified schematic of the experimental setup. Subsystems are identified by colour-coded labels.

The microdroplet diameter was measured with shadowgraphy. The total accuracy of the method was estimated to be ±100 nm. The main sources of uncertainty were the optical-system resolution and the blurring effect associated with microdroplet motion in the trap. The main uncertainty in the initial suspension concentration came from possible sedimentation of nanoparticles in the injector. Previous experiments [19,25] indicate that this uncertainty can be reduced either by minimizing suspension residence time in the injector or by matching the densities of the suspended and liquid phases. Here we used $TiO_2$ nanoparticles, which combine moderate density with high refractive index, and allowed a longer residence time to enable efficient acquisition of long droplet-evolution series. This choice reduced the accuracy of the initial concentration estimate, but was sufficient for testing the proposed classification concept.

In these experiments, we recorded speckle patterns generated by a 514 nm (green) diode laser beam of ~100 mW power and p-polarization. The optical path between the droplet and the CCD sensor was in free space (no additional lenses) to suppress static speckle that may originate from optical surfaces on the way. A single diaphragm, subtending a solid angle matched to the sensor area, was used to reduce the influence of stray light.

For the microdroplet experiments, we used $TiO_2$ nanoparticle suspensions in diethylene glycol (DEG). This ensured a high refractive-index contrast and, consequently, good speckle visibility. For $\lambda$ = 514 nm and $T$ = 22°C, $n_{TiO2} \cong 2.55$ for the amorphous material relevant to our NPs [26], whereas $n_{DEG} \cong 1.45$ [27]. The commercially available aqueous $TiO_2$ NP suspensions (by Corpuscular Inc.) were dried and weighed to determine the solids content independently, and then re-dispersed in DEG by several hours of sonication to obtain a mass concentration of 20 mg/mL. Lower concentrations were obtained by successive dilution of this base suspension. We used NPs of three distinctly different diameters: 100, 250 and 500 nm (C-TIO-0.10, Lot TM102; C-TIO-0.25, Lot TM307 and C-TIO-0.50, Lot TM012 respectively).

The exposure time was 20 ms at ~50 fps. It was shorter than the shortest measured speckle correlation time, estimated from DLS autocorrelation functions as ~30 ms for 100-nm $TiO_2$ particles at 20 mg/mL (Figure 3 – the source CSV file can be found at [28]; compare [29]). The correlation between consecutive speckle images was ~0.5, indicating substantial decorrelation within the recorded series.

In free-space speckle formation, the average speckle grain size $\delta x$ is determined by the relation [3]: $\delta x = \lambda L/D,$ where $\lambda$ is the light wavelength, $L$ is the propagation distance, and $D$ is the source diameter. For the geometry of the discussed experiment ($\lambda$ = 514 nm, distance between the microdroplet and the sensor $L$ = 47 mm, average microdroplet diameter $D$ = 100 μm), $\delta x \approx 240$ μm was much larger than the 7.4 μm pixel size of the KAI-0340 sensor in PIKE F-032 (Allied Vision) camera, used for speckle image recording. The experiment was performed beyond the characteristic distance $D^2/\lambda \approx 20$ mm.

In each experimental run, corresponding to a single-droplet observation, a sequence of 1000 48-bit RGB images was recorded. Only the 16-bit green channel was used. The background light, recorded without a droplet in the trap from a series of 100 images, was averaged and subtracted from each frame. The resulting images were filtered with a 2-D Gaussian kernel with a standard deviation of 2 pixels and then each image was individually min–max normalised. This reduced sensitivity to variations in absolute image brightness caused by factors such as the instantaneous droplet position in the illumination beam or fluctuations in beam power. The pre-processed images were stored as single-channel arrays. Altogether, ~$1.5\times10^6$ speckle images were collected.

Datasets for the selected one-, two-, and three-parameter classification tasks were constructed from these images. The arrays were loaded in MATLAB using an imageDatastore, with class labels assigned from folder names. For each task, the number of classes determined from the datastore was used to set the number of neurons in the final fully connected layer; the feature-extraction part of the network remained unchanged. For training control, the datastore was internally split by class into training, validation and internal hold-out subsets (0.8, 0.1 and 0.1, respectively). Because randomly selected frames from speckle-image data can remain partially correlated, this internal split was not used as the final measure of generalization.

Generalization to experimentally independent droplet series was assessed using test sets constructed separately from complete image series excluded from training. These included series acquired on experimental days not represented in the training set, as well as series from days represented in the training set only by other droplets. The setup was fully restarted for each experimental day. Each complete image series was also screened for visible experimental artefacts; series containing or suspected of containing glare, stationary speckle or other artefacts were excluded from training. The composition of the training/validation/internal-hold-out and independent test datasets for the three-parameter task is summarized in Table S2 and Table **S3** in the Supplementary material below. Sample images are available in [28].

The CNN architecture shown in Figure 2 was developed from the architecture used in our previous speckle-classification studies [15–17], and its code is available in [28]. It takes a 480×640×1 normalised speckle image as input and it is a residual multi-branch texture classifier consisting of an initial common feature-extraction stage followed by two residual skip-connected blocks and two parallel convolutional branches. The first branch contains convolutional layers with 128 and 256 filters, whereas the second contains layers with 256 and 384 filters. Their outputs are concatenated and passed through global average pooling, dropout, a fully connected layer, softmax activation and the final classification layer. The two-branch design was introduced to provide parallel feature representations with different capacities. During the three-parameter experiments, several variants of the branch part of the network architecture were explored using the same experimentally independent test set. None yielded a clear and consistent improvement, and the architecture described above was therefore retained. Thus, the test set remained independent of the training data, although it was also used for a limited comparison of architectural variants.

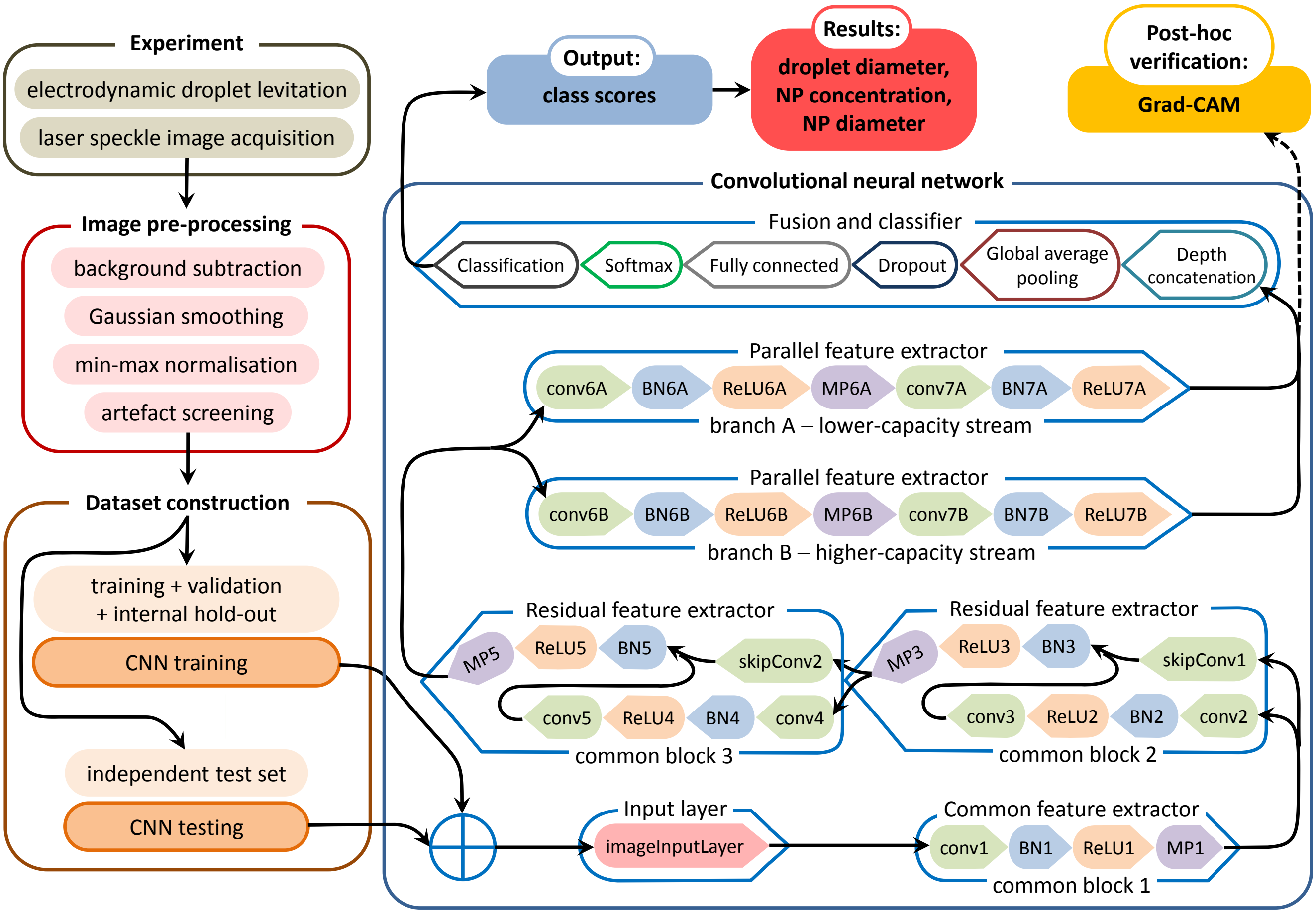

Figure 2 Image-analysis workflow and residual multi-branch CNN used for classification of speckle images. Abbreviations: conv, BN, ReLU and MP denote convolutional, batch-normalization, rectified-linear-unit and max-pooling layers, respectively.

ChatGPT (OpenAI) was used interactively during CNN development to assist with refinement of the code and exploration of training hyperparameters. All code modifications, parameter choices and analyses were reviewed and verified by the authors.

The CNN was implemented in MATLAB 2023a and trained using stochastic gradient descent with momentum (SGDM), with multi-GPU execution on two NVIDIA GeForce RTX 3090 GPUs. The complete set of training hyperparameters is provided in Table S1 in the Supplementary Material below.

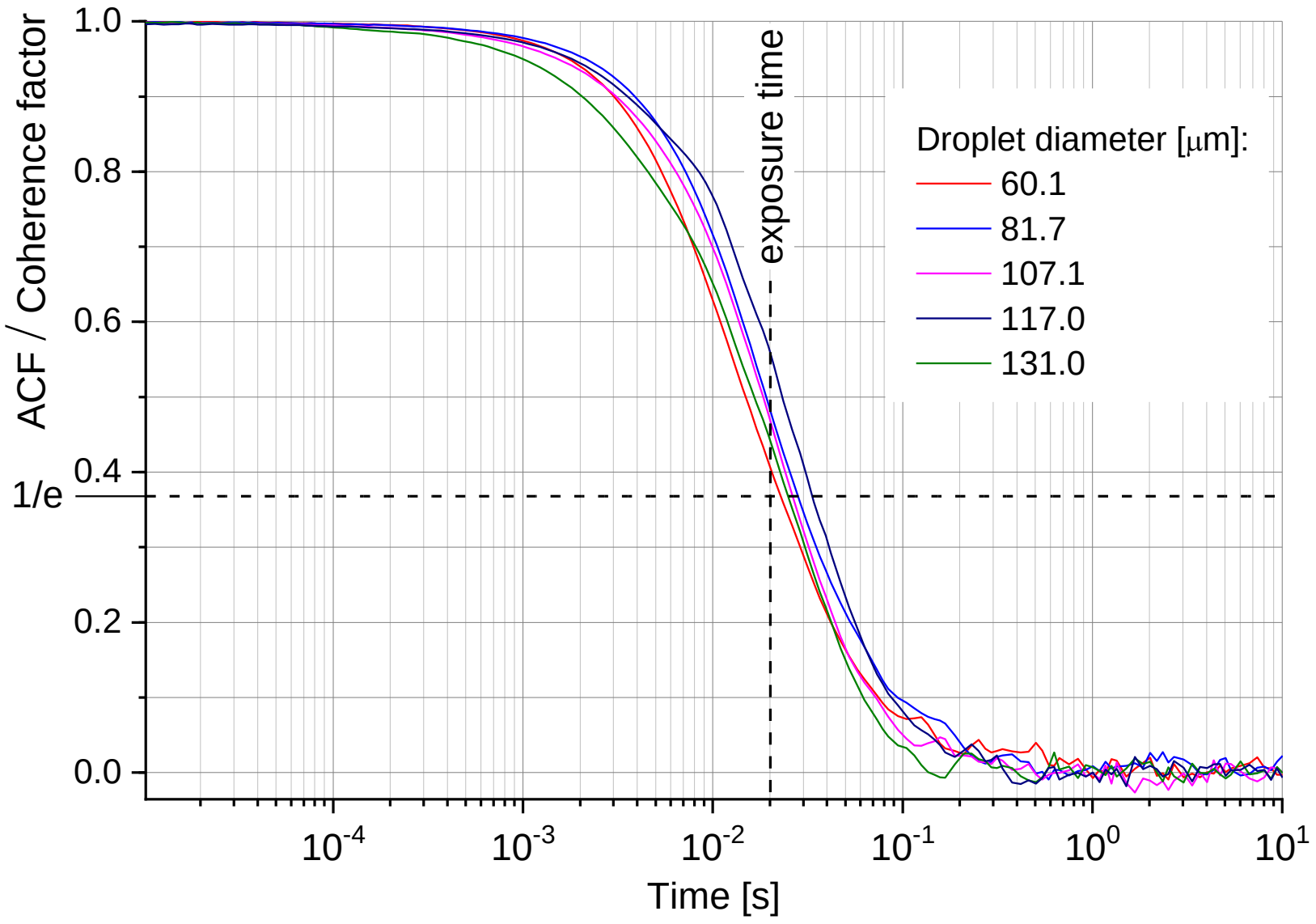

Figure 3 Autocorrelation function from dynamic light scattering on microdroplets of different diameters (colour-coded) at 20 mg/mL concentration of 100-nm diameter $TiO_2$ NPs. Exposure time and 1/e value are marked with the dashed lines.

For each image frame, the CNN returned a softmax score for each output class. For each 1000-frame series, the scores corresponding to each output class were summed over all frames. The resulting cumulative class scores were used to construct the confusion matrices. Since the softmax scores for each frame sum to unity, the cumulative class scores sum to the number of frames in the series, i.e. 1000 in the present experiments.

After training, Grad-CAM maps were inspected for representative images and selected artefact-suspected series as a qualitative check that the network response was distributed over speckle-texture regions rather than dominated by isolated image artefacts.

# Results

## Single-parameter determination of dense suspension microdroplets

### Recognizing diameter of a microdroplet of dense suspension

Unlike in [17], here we trained the CNNs on experimentally recorded speckle images from microdroplets. It turned out that, for a given suspension concentration and NP diameter, the diameter of a microdroplet could be assessed relatively easily using a trained CNN.

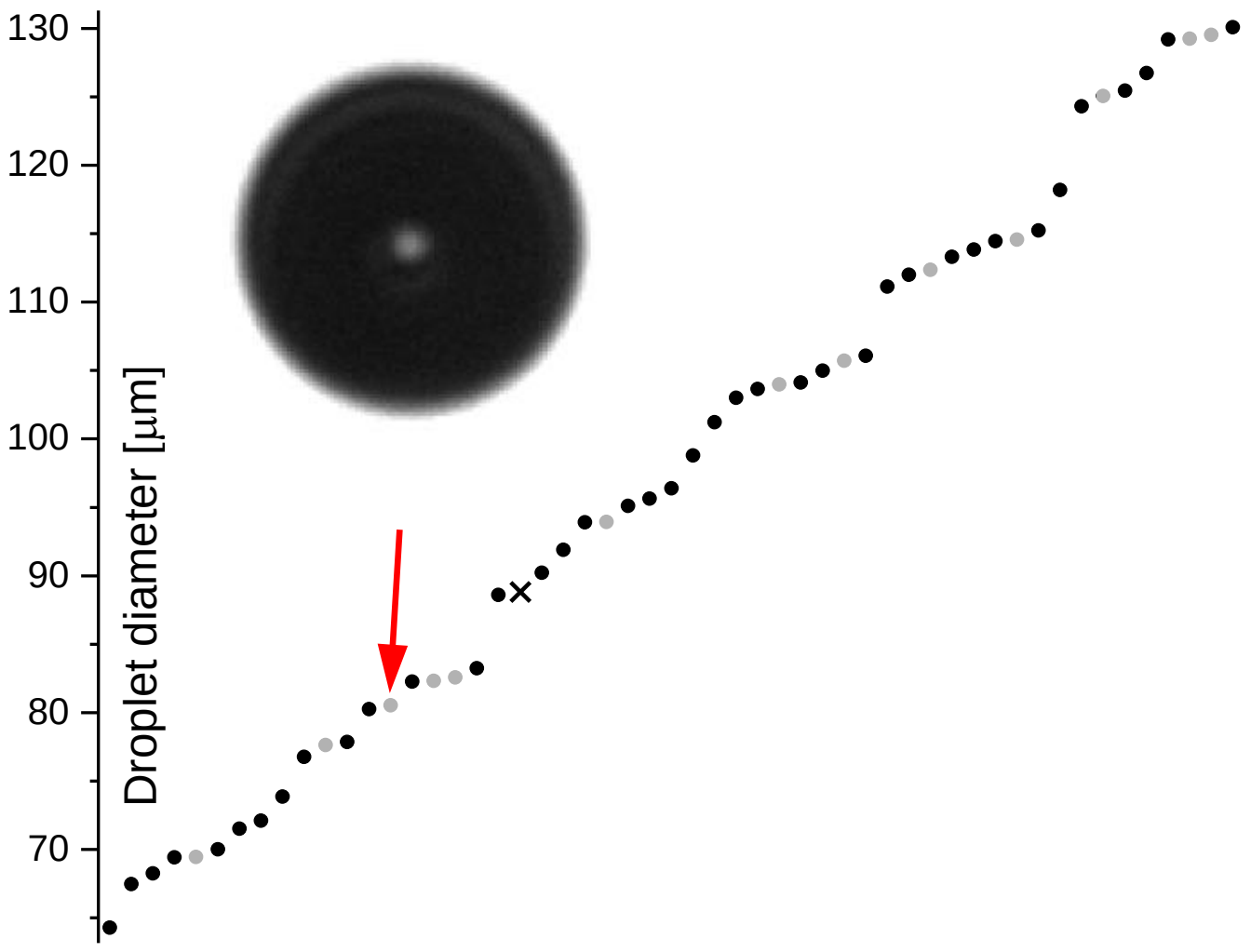

Figure 4 An example series of diameters of microdroplets with 20 mg/mL concentration of 100-nm diameter $TiO_2$ NPs, measured with shadowgraphy. The inset shows a recorded shadow of an 80.5-μm-diameter microdroplet indicated in the series with the red arrow. Speckle images corresponding to grey circles were excluded from training and used as the independent test set only. The experimental point marked with a cross was excluded from training and testing due to an artefact found in the corresponding images.

In Figure 4 we show a series of diameters – obtained from shadowgraphy – of microdroplets with 20 mg/mL concentration of 100-nm diameter $TiO_2$ NPs. An example shadowgraphy image is presented there in the inset. It corresponds to an 80.5-μm-diameter microdroplet indicated in the series with the red arrow. A central bright spot – the focal point at the surface of the sphere (droplet) visible in the shadowgraph indicates that nanoparticle scattering was weak enough for the droplets to retain an effectively droplet-like optical response, even at the highest mass concentration.

Because we did not have precise control of the initial injected droplet diameter, the sampled diameters were not evenly spaced, and some diameter ranges were overrepresented. To ensure a reasonably even diameter distribution for CNN training, we excluded "surplus" speckle images – corresponding to grey circles in Figure 4 – from training and used them to construct an independent data set instead. If an artefact was spotted by the experimenter in the collected images (glare, stationary speckle, etc.), the image series was excluded from training and from the main independent test sets. Additional checks on artefact-suspected image series gave poorer but still physically consistent classifications, supporting the conclusion that the network generalized well, and was not dominated by isolated artefacts.

In Figure 5 we show the cumulative classification-score matrix for the test set defined in Figure 4 (grey circles). The cumulative class scores are colour-coded, revealing some spread over neighbouring diameter classes. The cumulative class scores were used as weights to calculate the expected diameter and its standard deviation, as shown in Figure 6. Standard deviations are shown as vertical error bars and the relative uncertainty figures are shown in blue. It can be seen that the expected diameter values lie within the error bounds of the true values (red dashed Predicted = True line).

The Grad-CAM map in the inset in Figure 6 shows that the classification relied on irregular regions of the speckle image rather than on obvious static artefacts. The diameter accuracy estimated from the independent test set was better than ~6%.

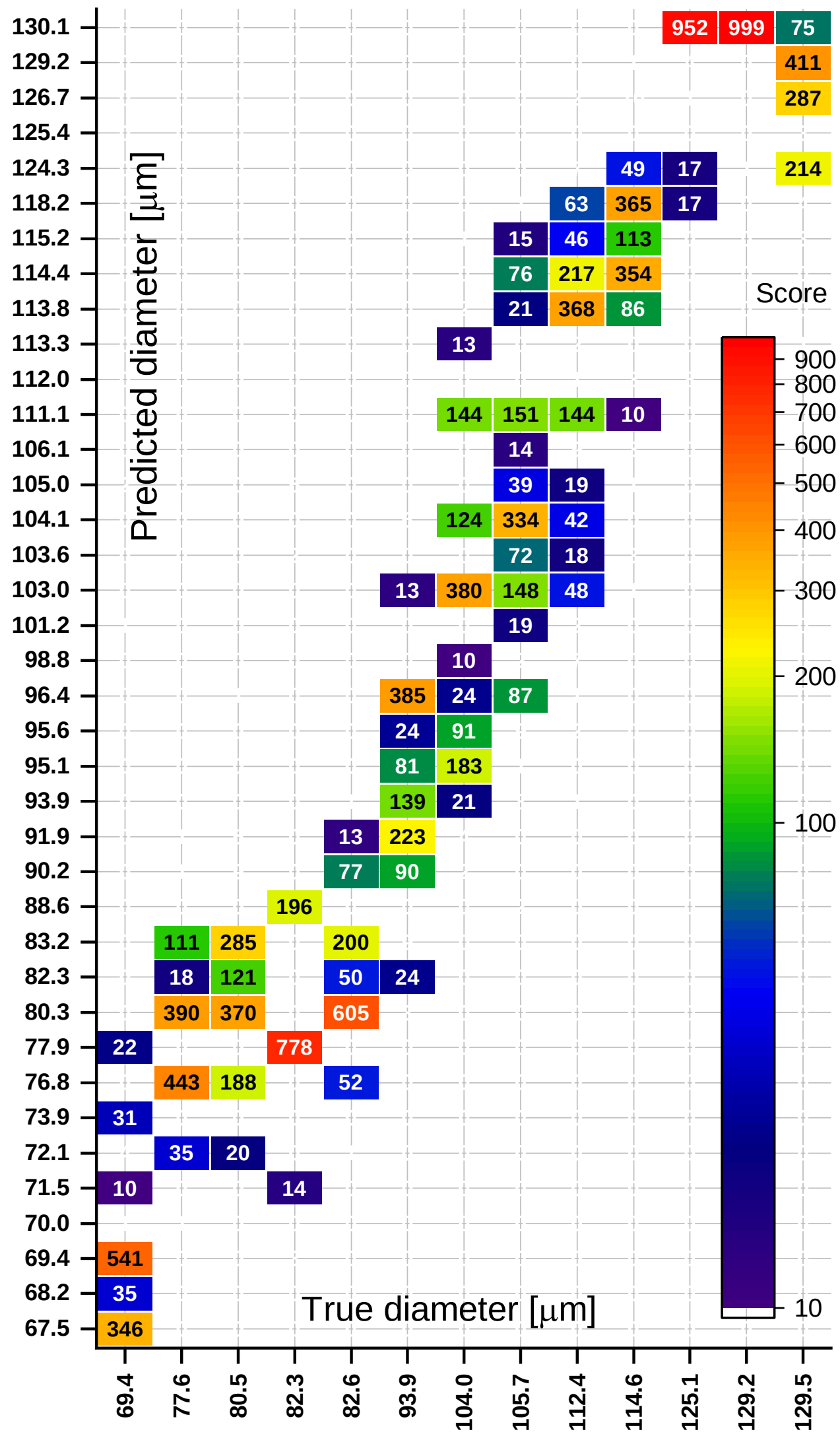

Figure 5 Cumulative classification-score matrix for the independent test set defined with grey circles in Figure 4. Diameter values are unique labels identifying each experimental run, thus their spacing is uneven. The cumulative class score colour scale is logarithmic for better visibility. Scores below 10 were hidden for clarity.

## Recognizing suspension concentration in microdroplets

Classifying NP concentration turned out to be significantly more demanding. Droplet diameter manifests itself strongly in the vertical spatial frequency of the speckles. It can be vividly illustrated by averaging a sequence of speckle images (as mentioned above) – resulting in characteristic regular fringes – see bottom inset in Figure 6. By contrast, suspension concentration manifests in some finer speckle image features, which are not really distinguishable to a human observer. First, we wanted to classify NP mass concentration for a fixed droplet diameter.

Because droplets of exactly the same diameter could not be reproduced for different concentrations, concentration classes were constructed from droplets within a selected diameter range. Narrow ranges of ±0.5 µm did not give robust classification, indicating strong sensitivity to diameter; therefore, a ±3 µm range was used. Then we tested the trained CNN on three different droplets from the selected range of diameters. An example result is shown in the left panel of Figure 7 for the droplet diameter range of 65±3 µm. We found that unambiguous clas-sification required adjacent concentration classes to differ by a factor of at least ~4. This requirement is consistent with the uncertainty of the initial concentration estimate, which broadens the effective concentration range within each class. This could be further improved by training a CNN on suspensions with better-defined concentrations.

### Recognition of suspension type as defined by nanoparticle size in microdroplets

In a complementary experiment, we investigated the classification of microdroplets containing NPs of the same material but of different diameters. In this case, it is more difficult to set parameters other than the NP material and the range of the microdroplet diameters as fixed parameters. Keeping the particle number constant across 100- and 500-nm particles would require a 125-fold difference in mass concentration and would also strongly change speckle brightness. We therefore performed NP-size classification at fixed mass concentration, recognizing that this couples particle diameter to particle number. To obtain a better generalization, the procedure of composing classes from microdroplet-diameter ranges was somewhat extended: image series corresponding to five fairly distant diameters from the ±4 μm range were combined, and four different droplets from the same diameter range were used for construction of the independent set. In the right panel of Figure 7, we present tests with microdroplets from the range of diameters of 100±4 μm, containing 5 mg/mL of $TiO_2$ NPs of either 100-, 250- or 500-nm diameter. The independent test set was classified unambiguously.

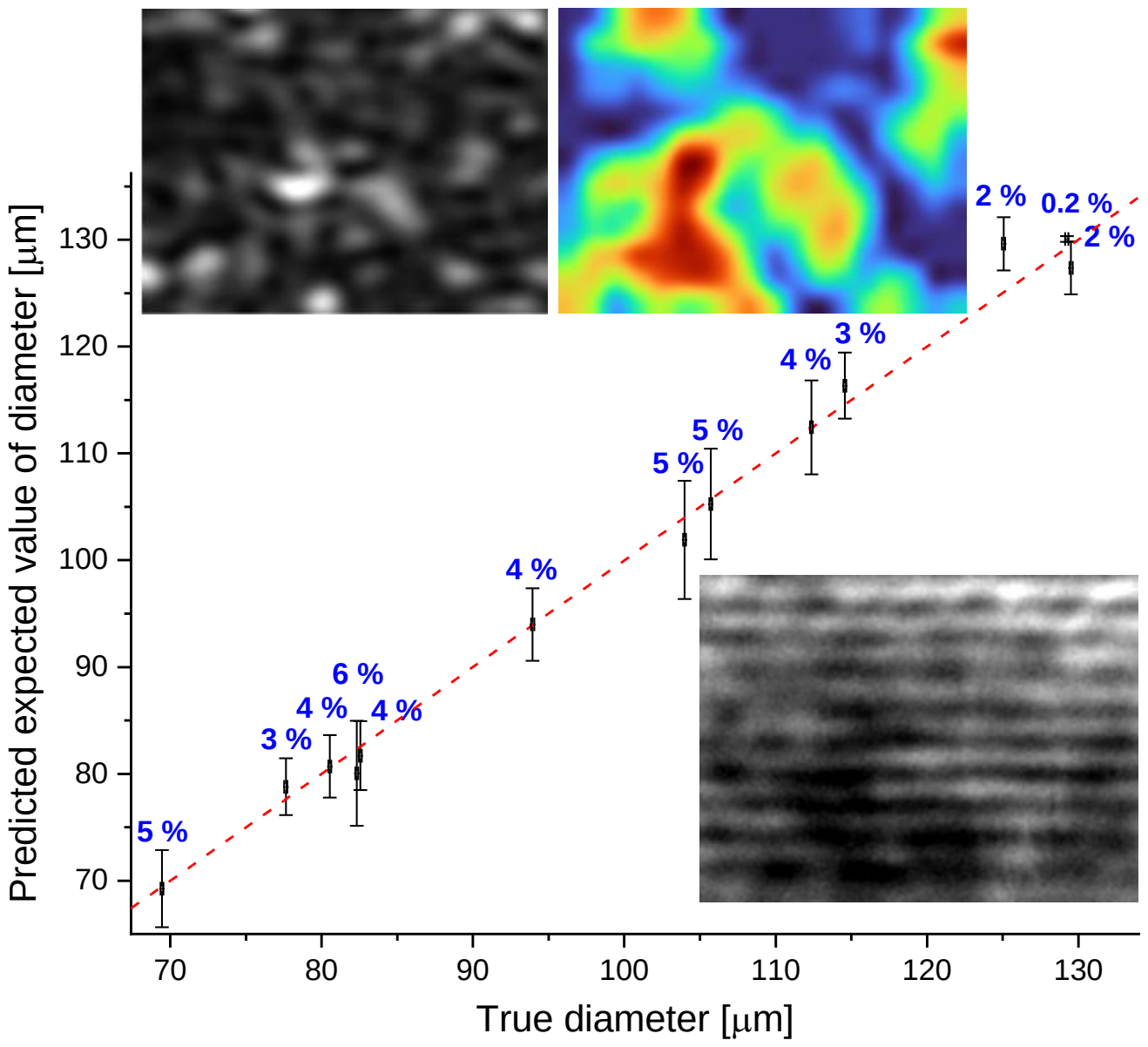

Figure 6 Expected values of diameters found with the classification of speckle images using our CNN for the droplet series from Figure 4. Vertical error bars correspond to standard deviation of classification, while horizontal error bars (significantly smaller) represent the accuracy of the shadowgraphy method that we used. The red dashed line corresponds to *Predicted* = *True*. The top-left inset shows a (filtered) speckle image – single (500[th]) frame – corresponding to the droplet shadow from the inset in Figure 4, while the top-right inset – a Grad-CAM map corresponding to this frame (colour scale is arbitrary here). Bottom inset shows an average speckle image (over 1000 frames) for this droplet.

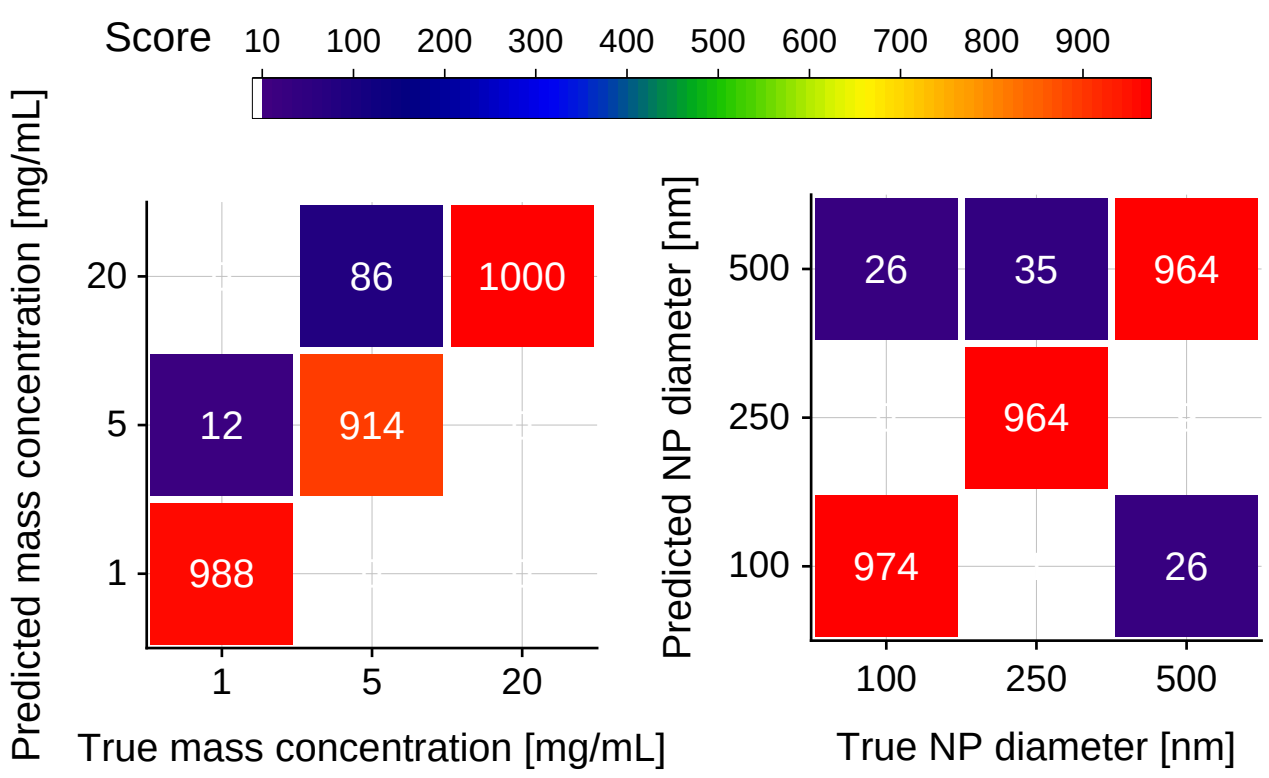

Figure 7 Confusion matrices for the independent single-parameter test sets corresponding to: three suspension concentrations for the microdroplet-diameter range of 65±3 μm – left panel, and three NP diameters for the NP mass concentration of 5 mg/mL and microdroplet-diameter range of 100±4 μm. The cumulative class score colour scale is linear. Scores below 10 were hidden for clarity.

## Simultaneous multi-parameter classification of suspension microdroplets

As an intermediate step toward multi-parameter recognition, we first trained CNNs for two 9-class tasks: nanoparticle concentration versus microdroplet-diameter range, and nanoparticle concentration versus nanoparticle diameter. In both cases, the independent test sets were classified unambiguously – only weak off-diagonal responses appeared for classes sharing one of the parameters. These tests confirmed that the network could use speckle-image features associated with more than one physical variable. Because the two-parameter tests served mainly as an intermediate validation step, their detailed confusion matrices are provided in the Supplementary material below – Figure S1 and Figure **S2**.

We then performed the final in-distribution classifica-tion task, in which the CNN was trained to classify suspension microdroplets simultaneously with respect to three parameters: nanoparticle mass concentration, microdroplet-diameter range and nanoparticle diameter. The classes were defined by three concentrations, 1, 5 and 20 mg/mL; three nominal microdroplet-diameter ranges centred at 73, 86 and 100 µm, each spanning approximately ±5 µm; and three nanoparticle diameters, 100, 250 and 500 nm, giving 27 classes in total. The results are shown in Figure 8. As in the two-parameter tests, weak cross-class "echoes" were observed, mostly between classes sharing one or two parameters. Nevertheless, the dominant responses remained associated with the correct classes, demonstrating that CNN-based analysis of speckle images can support simultaneous classification of multiple microdroplet and suspension parameters.

As an additional exploratory test, we examined whether the CNN trained for the three-parameter classification of single-component suspensions could generalize to more complex, dual-component suspensions. To this end, the trained network was exposed to speckle images generated by suspensions obtained by mixing single-component suspensions of the same nanoparticle mass concentration: 1, 5, and 20 mg/mL. This mixing protocol was chosen as a practical compromise, although it leads to markedly different number concentrations of the two nanoparticle fractions, especially for the maximally different particle diameters selected here, 100 and 500 nm. Mixing at equal number concentration would, in turn, strongly favour the larger particles because of their much stronger scattering response.

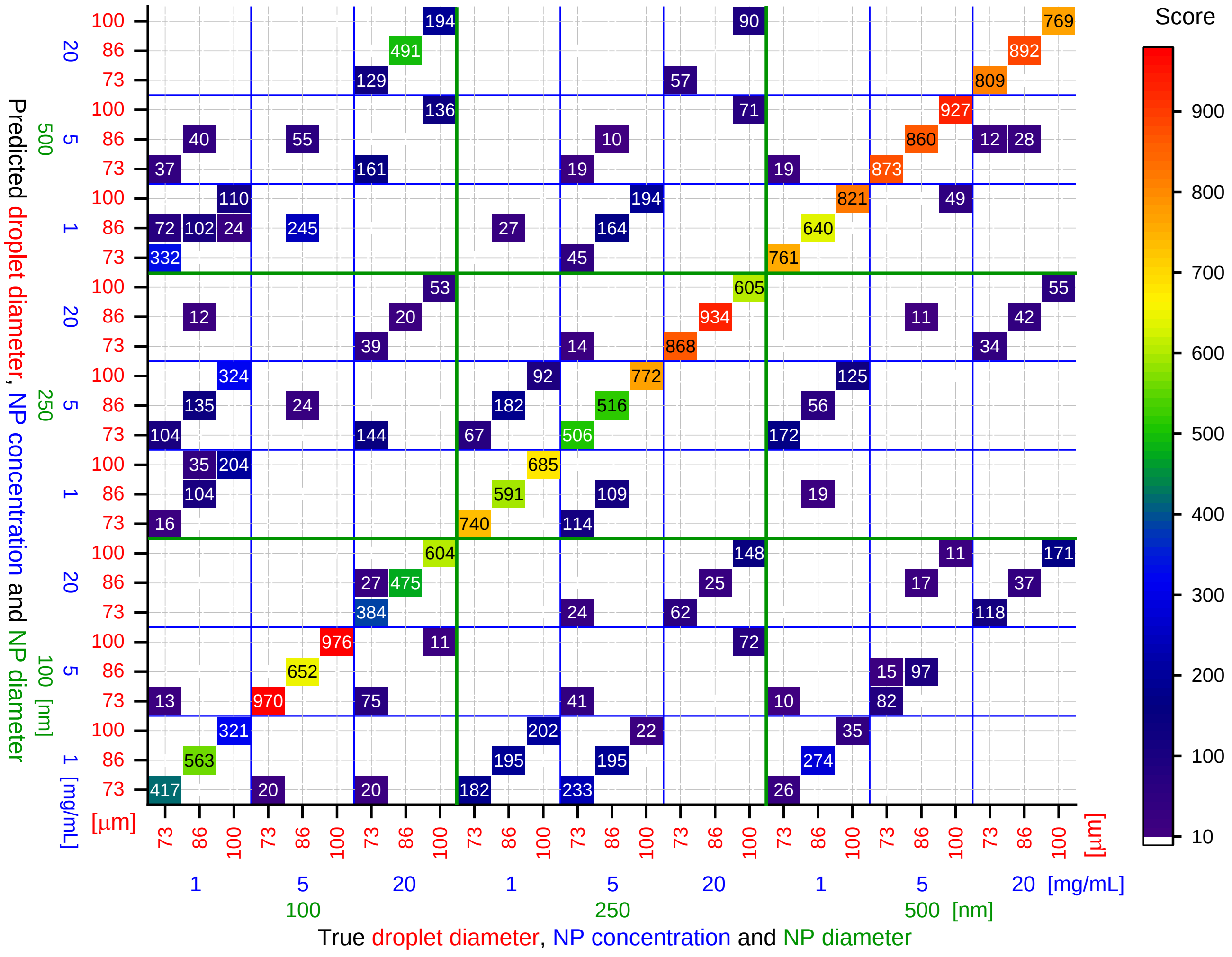

Figure 8 Confusion matrix for the test set consisting of 27 classes: three NP diameters (100, 250 and 500 nm), three suspension concentrations (1, 5 and 20 mg/mL) for each NP diameter, three nominal micro-droplet-diameter ranges centred at 73, 86 and 100 µm, each spanning approximately ±5 µm, for each NP concentration and NP diameter. Regions outlined by blue and green solid lines group classes sharing the same NP concentration and NP diameter, respectively, on both the predicted and true axes. The cumulative class score colour scale is linear. Scores below 10 were hidden for clarity.

For consistency with the previous experiments, three slightly narrower ranges of microdroplet diameters (73±4, 86±4, and 100±4 µm) were selected. The results for 1 and 20 mg/mL are shown in Figure 9. The result for 5 mg/mL was intermediate between these two cases and is therefore not shown for clarity. At the lower concentration of 1 mg/mL, the classification should rather be regarded as unsuccessful: only the microdroplet diameter was recognized with some accuracy, consistent with the strong manifestation of this parameter already discussed above, whereas the nanoparticle-size output tended toward an intermediate, effectively averaged class, with some accompanying uncertainty in concentration. The results improved with increasing mass concentration, and for 20 mg/mL they may be regarded as promising, although not yet fully satisfactory: both particle fractions became distinctly visible in the output, even though the tendency toward an intermediate class persisted. The concentration and microdroplet diameter were recognized unambiguously.

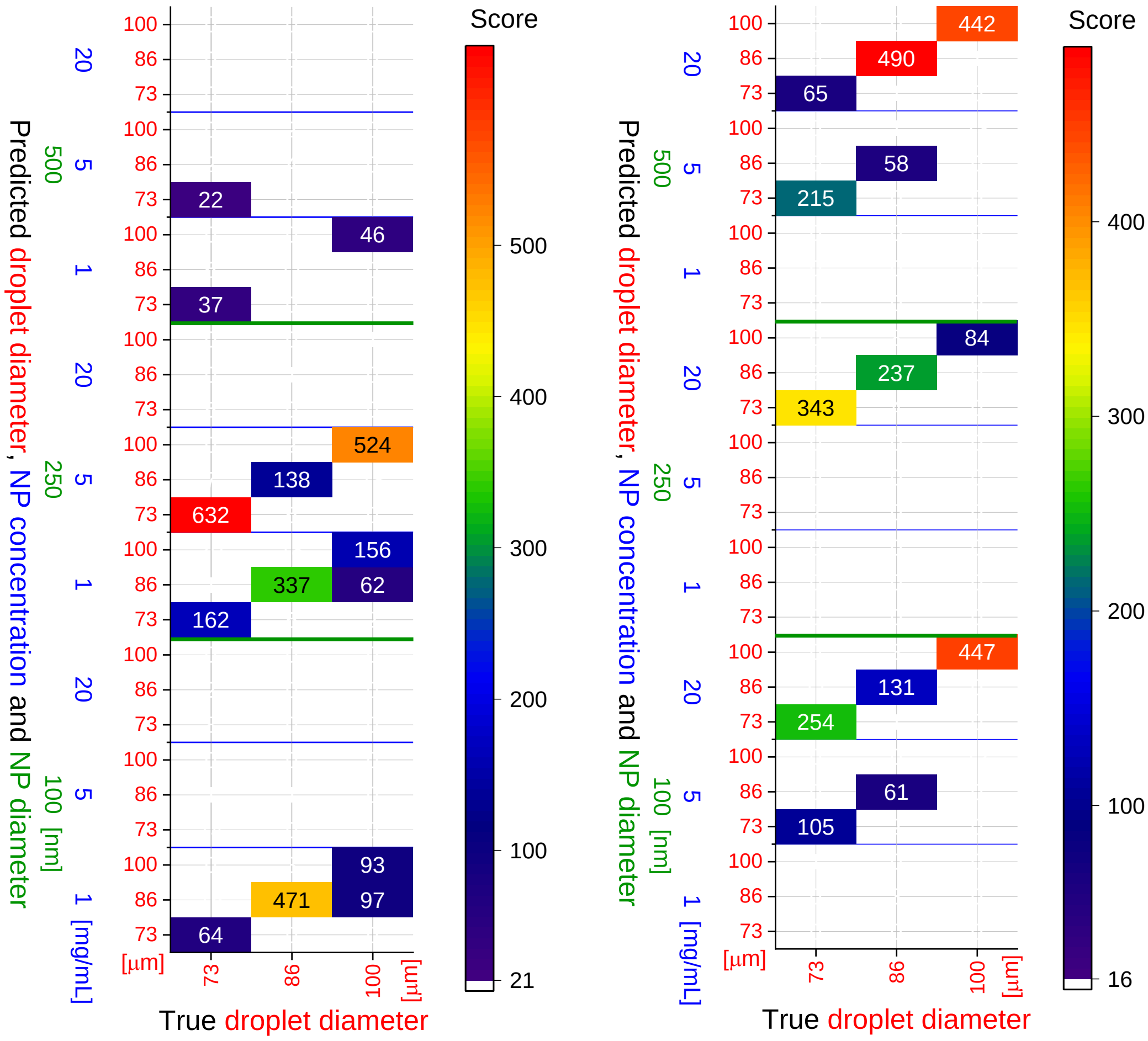

Figure 9 Confusion matrices for the independent test sets corresponding to dual-component suspensions composed of two NP diameter classes, 100 and 500 nm. For consistency with the previous experiments, three microdroplet-diameter ranges were selected: 73±4, 86±4, and 100±4 µm. The left and right panels correspond to suspension mass concentrations of 1 and 20 mg/mL, respectively. Regions outlined with blue and green solid lines correspond to the same predicted NP concentration and NP diameter, respectively. The cumulative class score colour scale is linear. Scores below the lowest values indicated in the colour scales were hidden for clarity.

These results indicate that the trained network retains some predictive value outside the single-component training distribution, but reliable decomposition of mixed nanoparticle populations will require dedicated training strategies or sequence-based analysis.

## Conclusions

This study demonstrates that laser speckle images recorded from single levitated suspension microdroplets contain sufficient information for CNN-based recognition of selected droplet and suspension parameters. For $TiO_2$ nanoparticle suspensions in diethylene glycol, the method enabled classification of droplet diameter, nanoparticle mass concentration and nanoparticle diameter, and, in the final in-distribution experiment, simultaneous three-parameter classification over 27 classes was achieved.

Among the tested quantities, droplet diameter was recognized most robustly, consistent with its strong manifestation in the spatial structure of the speckle patterns. Nanoparticle concentration was more demanding, most likely because its influence is encoded in subtler image features and because the initial concentration was subject to experimental uncertainty; nevertheless, reliable discrimination was obtained when concentration classes were sufficiently separated. Classes differing in nanoparticle diameter were also recognized, although the present fixed-mass-concentration design does not fully separate particle-size effects from accompanying changes in particle number.

The exploratory dual-component tests showed that a network trained on single-component suspensions can retain meaningful predictive capability outside the training distribution, especially at higher concentration, where both particle-size fractions became visible in the output. At the same time, the persistence of intermediate responses shows that the present approach is not yet a reliable method for decomposing mixed nanoparticle populations. Further improvements in concentration control, dataset design, network architecture and possibly sequence-based analysis should improve robustness and generalization, supporting the development of speckle-based optical diagnostics for more complex suspension microdroplets and aerosol-like systems.

## Acknowledgements

This research was funded in whole or in part by the National Science Centre, Poland, grant 2021/41/B/ST3/00069. For the purpose of Open Access, the author has applied a CC-BY public copyright license to any Author Accepted Manuscript (AAM) version arising from this submission.

## Declaration of generative AI and AI-assisted technologies in the manuscript preparation process

During the preparation of this work, the authors used ChatGPT (OpenAI) to assist with drafting the Abstract and Conclusions and improving the wording of several passages in other sections. After using this tool/service, the authors reviewed and edited the content as needed and take full responsibility for the content of the publication.

# **Supplementary material for:**
# Recognition of Nanoparticle and Microdroplet Parameters in Levitating Microdroplets of Suspension by Speckle Image Analysis Using Convolutional Neural Networks

*Yaroslav Shopa [1], Kwasi Nyandey [2,3], Daniel Jakubczyk [3,*]*

[1] Faculty of Mathematics and Natural Sciences. School of Exact Sciences, Cardinal Stefan Wyszynski University in Warsaw, Warsaw 01-815, Poland
[2] Department of Physics, School of Physical Sciences, Laser and Fibre Optics Centre, College of Agriculture and Natural Sciences, University of Cape Coast, Cape Coast, Ghana
[3] Institute of Physics, Polish Academy of Sciences, Warsaw 02-668, Poland
**Corresponding author: jakub@ifpan.edu.pl*

| Hyperparameter | |
|---|---|
| name | value |
| *Momentum* | 0.9 |
| *InitialLearnRate* | 0.005 |
| *MaxEpochs* | 18 |
| *Shuffle* | every-epoch |
| *ExecutionEnvironment* | multi-gpu |
| *ValidationPatience* | 4 |
| *LearnRateSchedule* | piecewise |
| *LearnRateDropPeriod* | 7 |
| *LearnRateDropFactor* | 0.1 |
| *L2Regularization* | 0.0001 |
| *GradientThreshold* | 1 |
| *MiniBatchSize* | 107 |
| *Dropout probability* | 0.3 |

Table S1 The hyperparameter values used for the final CNN training.

| Folder/label name | Image name: (consecutive image number)_(droplet diameter in μm) | | | | |
|---|---|---|---|---|---|
| 100nm_1mg_73um | *_68u2 | *_68u93 | *_71u52 | *_72u63 | *_74u56 |
| 100nm_1mg_86um | *_82u07 | *_83u89 | *_85u63 | *_90u04 | *_90u49 |
| 100nm_1mg_100um | *_95u77 | *_97u35 | *_99u59 | *_102u5 | *_105u28 |
| 100nm_5mg_73um | *_68u12 | *_68u64 | *_75u62 | *_76u27 | *_76u62 |
| 100nm_5mg_86um | *_82u08 | *_82u35 | *_85u19 | *_89u6 | *_89u8 |
| 100nm_5mg_100um | *_95u78 | *_96u46 | *_100u89 | *_101u63 | *_104u32 |
| 100nm_20mg_73um | *_68u24 | *_69u98 | *_73u68 | *_73u86 | *_75u92 |
| 100nm_20mg_86um | *_82u03 | *_82u27 | *_85u67 | *_89u56 | *_90u22 |
| 100nm_20mg_100um | *_96u07 | *_97u03 | *_100u41 | *_104u42 | *_104u45 |
| 250nm_1mg_73um | *_68u62 | *_69u7 | *_72u02 | *_74u92 | *_75u47 |
| 250nm_1mg_86um | *_83u1 | *_83u19 | *_85u59 | *_88u25 | *_88u47 |
| 250nm_1mg_100um | *_96u26 | *_96u79 | *_100u64 | *_103u25 | *_103u71 |
| 250nm_5mg_73um | *_67u89 | *_68u14 | *_70u9 | *_75u21 | *_76u24 |
| 250nm_5mg_86um | *_82u13 | *_82u58 | *_85u57 | *_90u35 | *_90u60 |
| 250nm_5mg_100um | *_96u08 | *_97u01 | *_98u28 | *_103 | *_103u57 |
| 250nm_20mg_73um | *_67u94 | *_68u31 | *_73u19 | *_75u84 | *_76u07 |
| 250nm_20mg_86um | *_82u2 | *_82u71 | *_85u89 | *_89u11 | *_90u04 |
| 250nm_20mg_100um | *_96u18 | *_96u44 | *_96u45 | *_100u71 | *_103u88 |
| 500nm_1mg_73um | *_68u56 | *_69u24 | *_73u66 | *_75u46 | *_76u14 |
| 500nm_1mg_86um | *_81u3 | *_81u64 | *_86u69 | *_90u25 | *_90u39 |
| 500nm_1mg_100um | *_95u51 | *_97u05 | *_100u64 | *_102u77 | *_103u79 |
| 500nm_5mg_73um | *_67u95 | *_67u97 | *_71u48 | *_73u77 | *_75u93 |
| 500nm_5mg_86um | *_82u95 | *_83u92 | *_85u74 | *_89u13 | *_89u96 |
| 500nm_5mg_100um | *_96u31 | *_97u02 | *_99u88 | *_101u03 | *_104u91 |
| 500nm_20mg_73um | *_68u39 | *_69u49 | *_75u88 | *_76u08 | *_76u17 |
| 500nm_20mg_86um | *_83u85 | *_84u24 | *_86u65 | *_88u22 | *_90u72 |
| 500nm_20mg_100um | *_98u78 | *_99u9 | *_100u49 | *_104u37 | *_104u52 |

Table S2 The training/validation/internal-hold-out dataset corresponding to the 3-parameter task. For each droplet of a given diameter, 1,000 images are available; the '*' symbol indicates the consecutive image number. The letter 'u' was used instead of the decimal point in automatically generated file and folder names containing droplet-diameter values.

| Image series group | | | Image name: (consecutive image number)_(droplet diameter in µm) | | | |
|---|---|---|---|---|---|---|
| NP diameter | NP concentration | droplet-diameter range | | | | |
| 100 nm | 1 mg/mL | 73 µm | *_70u17 | *_72u08 | *_73u49 | *_73u81 |
| | | 86 µm | *_84u24 | *_84u96 | *_86u51 | *_87u55 |
| | | 100 µm | *_98u95 | *_99u2 | *_99u61 | *_100u42 |
| | 5 mg/mL | 73 µm | *_73u57 | *_74u24 | *_74u50 | *_75u42 |
| | | 86 µm | *_85u93 | *_86u11 | *_86u49 | *_87u23 |
| | | 100 µm | *_97u41 | *_97u51 | *_101u46 | *_102u22 |
| | 20 mg/mL | 73 µm | *_72u11 | *_72u49 | *_73u04 | *_73u39 |
| | | 86 µm | *_83u24 | *_85u25 | *_87u47 | *_87u71 |
| | | 100 µm | *_99u73 | *_100u26 | *_100u55 | *_100u98 |
| 250 nm | 1 mg/mL | 73 µm | *_72u34 | *_72u45 | *_72u92 | *_73u28 |
| | | 86 µm | *_83u58 | *_85u22 | *_85u35 | *_86u47 |
| | | 100 µm | *_98u95 | *_99u86 | *_101u17 | *_101u88 |
| | 5 mg/mL | 73 µm | *_68u79 | *_68u81 | *_70u46 | *_70u85 |
| | | 86 µm | *_83u16 | *_84u6 | *_86u46 | *_88u26 |
| | | 100 µm | *_97u08 | *_100u16 | *_100u63 | *_102u35 |
| | 20 mg/mL | 73 µm | *_70u87 | *_71u36 | *_74u59 | *_75u03 |
| | | 86 µm | *_85u01 | *_86u33 | *_86u4 | *_86u76 |
| | | 100 µm | *_97u19 | *_99u88 | *_101u28 | *_105u71 |
| 500 nm | 1 mg/mL | 73 µm | *_70u01 | *_71u8 | *_72u54 | *_74u44 |
| | | 86 µm | *_82u95 | *_83u32 | *_85u98 | *_89u04 |
| | | 100 µm | *_97u77 | *_101u456 | *_101u460 | *_103u12 |
| | 5 mg/mL | 73 µm | *_70u11 | *_70u54 | *_71u88 | *_73u87 |
| | | 86 µm | *_84u8 | *_85u01 | *_86u47 | *_86u97 |
| | | 100 µm | *_98u22 | *_99u48 | *_99u82 | *_99u87 |
| | 20 mg/mL | 73 µm | *_73u72 | *_74u2 | *_72u05 | *_76u38 |
| | | 86 µm | *_84u48 | *_87u2 | *_86u5 | *_87u37 |
| | | 100 µm | *_99u74 | *_101u58 | *_101u95 | *_102u55 |

Table S3 The independent test data set corresponding to the 3-parameter task. For each droplet of a given diameter, 1,000 images are available; the ‘*’ symbol indicates the consecutive image number. The letter ‘u’ was used instead of the decimal point in automatically generated file and folder names containing droplet-diameter values.

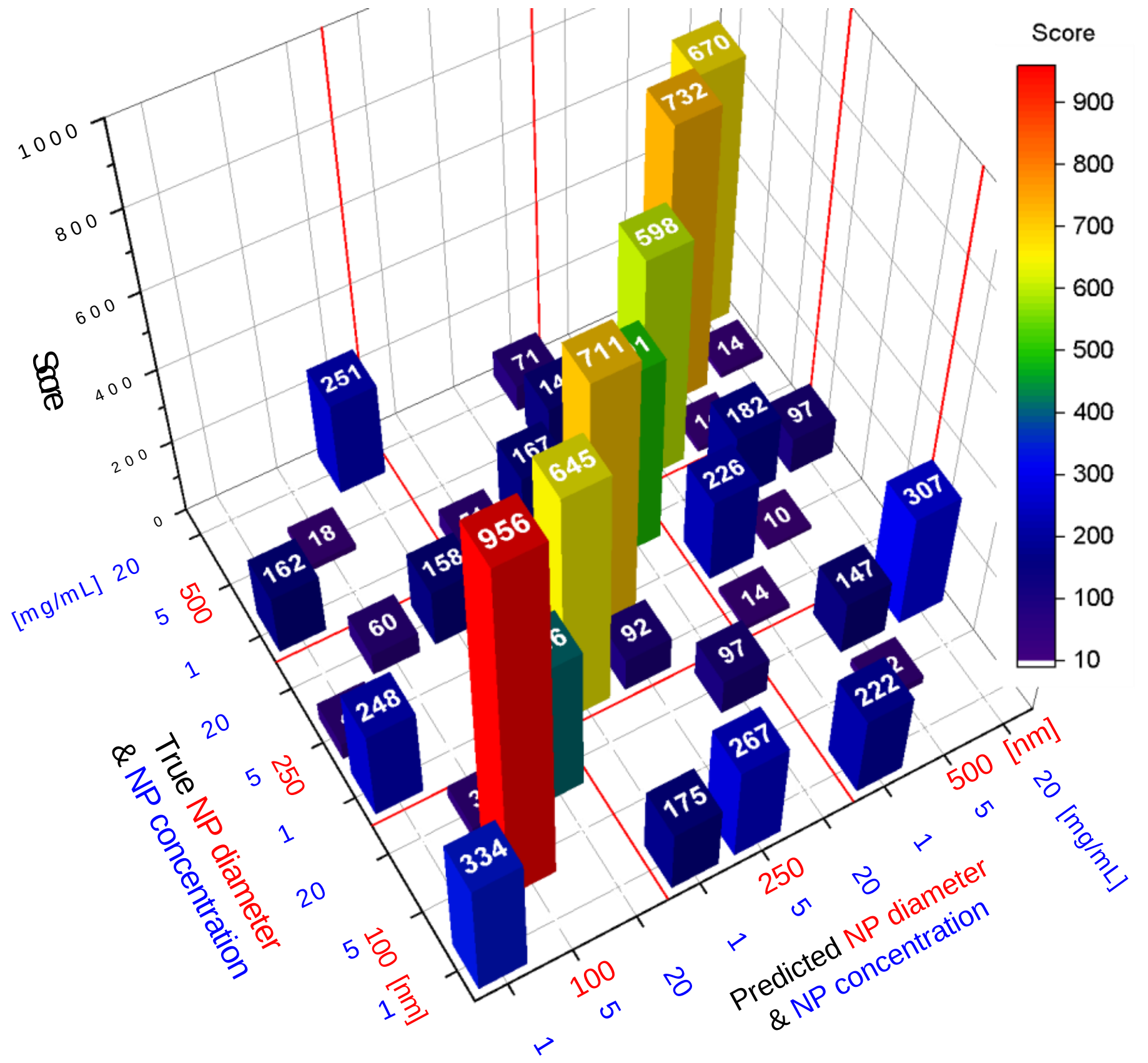

Figure S2 A 3D visualization of the confusion matrix for the test set consisting of 9 classes: three NP diameters (100, 250 and 500 nm) versus three suspension concentrations (1, 5 and 20 mg/mL) for each NP diameter. Microdroplet diameters were in the range of 65±3 μm. Regions outlined with red solid lines correspond to the same NP diameter, on both the predicted and true axes. Scores below 10 were hidden for clarity.

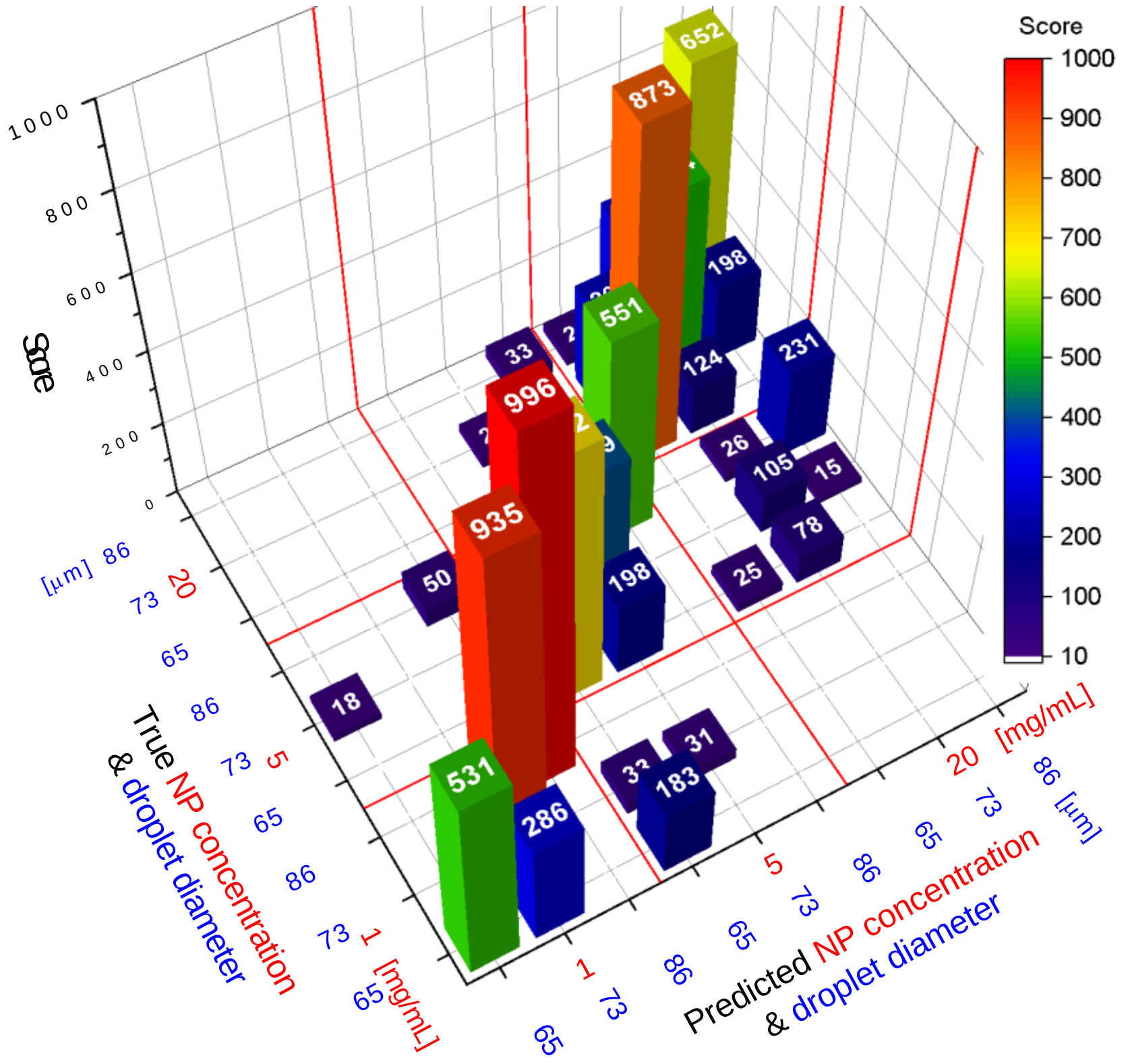

Figure S1 A 3D visualization of the confusion matrix for the test set consisting of 9 classes: three suspension concentrations (1, 5 and 20 mg/mL) versus three ranges of microdroplet diameters (65±4, 73±4 and 86±4 μm) for each NP concentration. Regions outlined with red solid lines correspond to the same concentration, on both the predicted and true axes. Scores below 10 were hidden for clarity.